%% file: CadutaCorpi.tex
\documentclass[later,twocolumn,11pt]{article}
\usepackage[spanish]{babel}
\usepackage{ae}
\usepackage[latin1]{inputenc} % Escritura en castellano con acentos
\usepackage[T1]{fontenc} % Escritura en castellano con acentos
\usepackage{layout}
\usepackage{array}
\usepackage{graphicx}
\usepackage{float}
\usepackage{amsmath}
\usepackage{multicol}
\usepackage{rotating}  %--Rota una palabra
\hyphenrules{nohyphenation} %--No parte palabras en español
\usepackage[superscript]{cite} %---Lista citas biliograficas continuas pero supra
\usepackage[all]{xy}
\usepackage{amsmath,tikz}
\usetikzlibrary{calc}
\usepackage[]{cite}   %---Lista citas biliograficas continuas per p.ej. [1-5]
\usepackage{anysize}  %--Controla las margenes
\marginsize{1.5cm}{1.5cm}{1.0cm}{2.8cm}
\usepackage{endnotes} %--control de notas
\begin{document}
\renewcommand{\tablename}{Tabella}
\renewcommand{\abstractname}{Abstract}
\renewcommand{\thefootnote}{\arabic{footnote}}
\renewcommand{\figurename}{Fig}

\title{
        {\LARGE  Caduta dei gravi nell'aria: uno studio sperimentale sull'evoluzione dell'apprendimento attraverso PI}\\
}
\author
{ 
Paco H. Talero L. $^{1,2}$\\
$^{1}$ {\small Universidad Central, Facultad de Ingeniería y Ciencias Básicas}\\ 
\small Carrera 5 No 21-38, Bogot\'{a}, D.C.Colombia.\\
$^{2}$ {\small Universidad Jorge Tadeo Lozano,  Facultad  Ciencias Naturales e Ingeniería}\\
\small Carrera 4 No.22-61, Bogot\'{a}, D.C.Colombia.\\
}%-------------------------------------------------------------------------------------------
\date{}

\twocolumn
[
\begin{@twocolumnfalse}

\maketitle %--El titulo queda en una sola columna
\begin{abstract}
Nel campo della fisica educativa, le metodologie attive sono state  più efficienti delle metodologie 
tradizionali. Ci sono  due metodologie molto efficienti,  PODS  e  PI. Nonostante il grande successo di 
queste metodologie, ci sono tre problemi metodologici che ne offuscano la loro efficacia: primo,  il  pre e pos-test  non
permette di conoscere che delle cose siano  successe  attraverso il processo di apprendimento; secondo,  le risposte che gli studenti normali 
danno alle domande del test  stanno  influenzate per le loro interazioni con studenti di livello piú 
superiore;  terzo, tradizionalmente nella letteratura si mostrano studi privi di esperimenti. In questo 
lavoro si rapporta un esperimento sociale di apprendimento attraverso PI con $42$ studenti su  un piccolo 
concetto fisico (Caduta gravi nell'aria) in cui i processi non sono stati involucrati dai tre problemi 
sopra citati. Si è scoperto che l'apprendimento si è evoluto in modo più o meno  lineare senza picchi, discese 
e interferenze che  di solito compaiono nella letteratura.\\

\textbf{Parole chiave:} PI, fisica educativa,  Caduta dei gravi.\\ 

Physics Education Research has shown that the active methodologies are better than traditional methodologies, the most popular are PODS and PI. However, 
there are three problems that blur their effectiveness: first, the pre and post-test do not allow to know what happens in the learning process; second, 
the answers that average - performance students give to questions of a test are influenced by the answers of the more advanced students; third, in many 
types of research on PI the students do not experiment. In the present work, it is exposed a research on the learning through PI with $42$ students over a 
small physical concept (Bodies falling in air), the interaction between the students was designed to correct the aforementioned problems. It was found a 
linear evolution of the gain $G$, with high concentration in the correct model and it was not observed peaks or interference.\\

\end{abstract}
%\date{}
\end{@twocolumnfalse}
]

\section{Introduzione}
Circa quaranta anni fa alcuni dipartimenti di fisica, principalmente negli Stati Uniti, 
si sono focalizzati sull'attenzione dell'apprendimento della fisica di base  a livello universitario e 
hanno formato la Physics Education Research (PER) per fare una ricerca  in questo campo da un punto di vista 
scientifico \cite{Redish,Ciencia,Julio}. Nell'anno $2000$ alcuni ricercatori trattavano già in pratica la 
fisica educativa come una scienza empirica applicata \cite{Ciencia}.\\

Le ricerca della PER hanno permesso di capire, tra l'altro, che: le metodologie attive sono molto di più 
efficienti delle metodologie espositive tradizionali\cite{Hake}; i questionari (strumenti di indagine) su 
diverse tematiche sono utili per trovare problematiche di apprendimento \cite{Hestenes1,Beichner,BEMA}; i 
ricercatori hanno bisogno di partire dalla scienza cognitiva per definire gli scopi e le limitazioni 
presenti per natura negli studenti quando imparano \cite{RedishPC,grote}; e che, gli strumenti quantitativi 
per misurare l'impatto dell'istruzione ha bisogno di dare valore  alla relazione tra l'insegnamento e 
l'apprendimento \cite{Hake,Bao2,Evol}.\\

Eppure, negli ultimi anni, i ricercatori hanno osservato che quando si fa una ricerca con il pre e 
pos-test non si sa che cosa potrebbe  succedere nel processo, quindi i risultati di questi tipi di 
ricerca potrebbero essere  sbagliati. Anche l'influenza degli studenti avanzati sulle risposte degli 
studenti normali è stata osservata dai ricercatori, le quali, quindi,  sono poco credibili in alcuni 
risultati di ricerca effettuati con la metodologia Peer Instruction (PI).\\ 

In questo lavoro si è studiato l'apprendimento qualitativo di un piccolo concetto della caduta di 
corpi nell'aria attraverso la metodologia PI, si è fatta la correzione dei  due problemi precedenti 
e si sono proposti esperimenti diversi. L'ipotesi che è stata formulata era: se si fa un esperimento 
senza l'influenza di questi problemi, gli studenti avranno profitto elevate, come poi si è verificato. \\ 

Questo lavoro è così suddiviso: nella sezione $2$ si presenta lo stato dell'arte del campo di ricerca; 
nella sezione $3$ si illustra la metodologia proposta per sviluppare la ricerca; nella sezione $4$ si 
presentano i  risultati;  le discussioni  sono spiegate nella sezione $5$ e le conclusioni sono 
sviluppate nella sezione $6$.\\ 

\section{Lo stato dell'arte}

Questo lavoro è inquadrato nel campo della fisica educativa, esplora aspetti 
dell'evoluzione dell'apprendimento di un micro - contenuto fisico (MCF), utilizza
la metodologia basata su PI e si basa su alcuni principi cognitivi di apprendimento.
 
\subsection{Micro-contenuto fisico}

Un MCF è un argomento fisico molto breve, chiaramente definito, delimitato sia nel 
contenuto che nella profondità, la quale dipende dalla popolazione e dal contesto generale che si vuole 
insegnare.  Anche, un MCF è definito  da un comitato esperti che conoscono la disciplina, il contesto e sono
formati in pedagogica \cite{Evol}. Un MCF si materializza attraverso un insieme di attributi  concettuali 
denotati come $F_0$, $F_1$, $F_2,  \ldots ,F_N$  che a sua volta sono discriminati per ogni attributi 
concettuale $q_0$, $q_1$,$q_2$,...,$q_m$ i quali  sono scelti uguali in numero e valore per ogni attributo
con\-cettu\-ale. 
%-------hasta aquí cooregido por p vito 1
\subsection{Principi cognitivi di apprendimento}
L'importanza delle neuroscienze cognitive nell'apprendimento della fisica elementare è stata sempre presente 
nelle ricerche \cite{RedishPC},  in modo  approssimativo Redish e Grote  la riassume attraverso sei principi 
generali \cite{RedishPC,grote}:

\begin{enumerate}
	\item Uno studente che impara un nuovo argomento deve fare un rapporto tra la nuova conoscenza e la sua precedente.  
	\item L'elaborazione della conoscenza dello studente dipende in gran parte dal contesto.  
	\item Quando uno studente ha già uno schema sbagliato installato, fare un cambiamento importante sarà molto difficile.
	\item In generale gli studenti hanno  diversi modi di imparare. 
	\item In generale, gli studenti imparano più efficacemente attraverso le interazioni con altri studenti che solo con le istruzioni del professore. 
	\item L'istruzione a intervalli distanziati è più efficace dell'istruzione fatta in un unico blocco di lavoro.     
\end{enumerate}

\subsection{Istruzione tra studenti}
La metodologia PI è  nata  dall'iniziativa del professor Eric Mazur dell'Università Harvard
 \cite{10Anos,Mazur},  consiste nella seguente serie di passaggi:  

\begin{enumerate}
\item  Prima di andare a lezione gli studenti hanno già letto il materiale teorico di lavoro, il quale  è sviluppato  attraverso una 
        trasposizione didattica diretta precisamente a questa  popolazione.  
\item  Il professore fa  una domanda, che di solito sembra un enigma, ai suoi studenti  e li invita a risponderla individualmente.
 \item Il professore,  l'assistente oppure il sistema di acquisizione dati raccolgono le risposte.
\item  Il professore fa dei gruppi di due o tre studenti  e chiede una risposta univoca alla domanda precedente. 
\item  Il professore, l'assistente oppure il sistema di acquisizione dati raccolgono le risposte date dalle coppie di studenti.
\item  Il professore risolve il problema posto, risponde alle domande che gli studenti formulavano e dopo li invita alla riflessione. Poi si torna al punto $2$.  
\end{enumerate}

\subsection{Metriche di apprendimento}  
La PER ha progettato diverse metriche per quantificare l'apprendimento, le più utilizzate sono 
il fattore di Hake \cite{Hake} e l'indice di concentrazione di Bao \cite{Bao2}.\\ 

\subsubsection{Il fattore di Hake}  
Il Fattore Hake è una metrica che misura l'apprendimento di un gruppo di  studenti utilizzando un unico 
questionario come pre e post-test. Questo strumento è  denotato con la lettera  $G$,  calcola il cambiamento 
del punteggio dello studente ed è definito come:
\begin{equation}
G=\frac{ S_{f}-S_{i} }{ 1-S_{i} }.
\end{equation} 
Qui  $S_i$ è il punteggio del gruppo nel pre-test e $S_f$, nel post-test; i due punteggi sono standardizzati 
all'unità. $G$ è una misura dell'efficacia di una metodologia applicata a un  corso particolare di fisica. 
Il risultato è stato classificato in alto, medio  e basso. È alto quando $G \ge 0,7$, medio quando  
$0,3 \leq G <0,7$ e basso quando $G<0,3$. 

\subsubsection{L'indice di concentrazione}  
L'indice  di concentrazione $C$ rende conto della distribuzione delle risposte che $N_e$ studenti 
hanno a una domanda con  $m-1$  opzioni sbagliate e un'unica risposta corretta. Cioè, $C$ dice quanti 
studenti hanno risposto a ciascuna delle opzioni.  La concentrazione è definita come

\begin{equation}   %-dejar renglon produce error
C = \frac{\sqrt{m}}{\sqrt{m}-1}\left(\sqrt{\frac{\sum_{i=1}^{m}{n_i^2}}{N_e}}-\frac{1}{\sqrt{m}}       \right),
\end{equation}
dove $n_i$ è il numero di visite alla opzione $i$ di $m$. Per fare una corretta interpretazione è 
necessario correlare la concentrazione $C$ con il punteggio  $u$ e definire i suoi estremi di 
concentrazione minimo $Cimn$ e massimo $Cmas$.\\

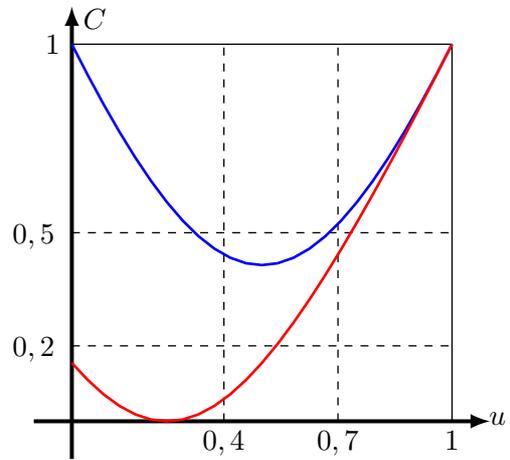
\begin{figure}[!htp]
\begin{center}
	\begin{tikzpicture}[domain=0:1,scale=5]
	\draw [-latex,black,line width=1.5pt] (-0.1cm,0cm) -- (1.1cm,0cm);%--Eje x----
	\draw [-latex,black,line width=1.5pt] (0cm,-0.1cm) -- (0cm,1.1cm);%--Eje y---
	\coordinate [label=below:\textcolor{black} {$u$}] (x) at  (1.12cm,0.05cm);%-x
	\coordinate [label=below:\textcolor{black} {$C$}] (x) at  (0.06cm,1.12cm);%-y
	\coordinate [label=below:\textcolor{black} {$1$}] (x) at  (1.0cm,0.0cm);%-x
	\coordinate [label=below:\textcolor{black} {$1$}] (x) at  (-0.05cm,1.05cm);%-y
	\coordinate [label=below:\textcolor{black} {$0,4$}] (x) at  (0.4cm,0.0cm);%-0,4
	\coordinate [label=below:\textcolor{black} {$0,7$}] (x) at  (0.7cm,0.0cm);%-0,4
	\coordinate [label=below:\textcolor{black} {$0,2$}] (x) at  (-0.1cm,0.25cm);%-0,2
	\coordinate [label=below:\textcolor{black} {$0,5$}] (x) at  (-0.1cm,0.55cm);%-0,2
	
     \draw [black,line width=0.5pt] (0cm,1.0cm) -- (1cm,1cm);%----
     \draw [black,line width=0.5pt] (1cm,1.0cm) -- (1cm,0cm);%----
     \draw [black,dashed,line width=0.5pt] (0cm,0.2cm) -- (1cm,0.2cm);%----
     \draw [black,dashed,line width=0.5pt] (0cm,0.5cm) -- (1cm,0.5cm);%----
     \draw [black,dashed,line width=0.5pt] (0.4cm,0.0cm) -- (0.4cm,1.0cm);%----
     \draw [black,dashed,line width=0.5pt] (0.7cm,0.0cm) -- (0.7cm,1.0cm);%----
     %-------------Funciones m=4--------------------
     \draw[color=blue,line width=1pt]   plot ({\x},{2.0*sqrt(1-2.0*\x+2.0*\x*\x)-1.0}); %\x-(\x*\x)/2
	\draw[color=red,line width=1pt]   plot ({\x},{2.0*sqrt(1.0/3.0-2.0*\x/3.0+4.0*\x*\x/3.0)-1.0});
	\end{tikzpicture}
\caption{Concentrazione contro punteggio.}
\label{ConC}
\end{center}
\end{figure}

Le espressioni ottenute sono:

%--------
\begin{equation} \label{ca3}  %-dejar renglon produce error
  C_{min}= \frac{\sqrt{m}}{\sqrt{m}-1} \left(  \sqrt{  \frac{(1-u)^{2}}{m-1}  +u^{2}}-\frac{1}{\sqrt{m}} \right)
\end{equation}
%--------
e
%------------
\begin{equation} \label{ca6}  %-dejar renglon produce error
  C_{mas}= \frac{\sqrt{m}}{\sqrt{m}-1} \left( \sqrt{(1-u)^{2} +u^{2}} -\frac{1}{\sqrt{m}} \right). 
\end{equation}
%-----
Nella Fig.\ref{ConC} sono stati mostrati in rosso e blu $Cmas$ e $Cmin$, rispettivamente. 
Queste curve definiscono le possibili zone  in cui può trovarsi il risultato di una prova, dei punti 
fuori di  questa zona sono privi di significato.
%-----
%------------------

Il diagramma di concentrazione mostrato nella Fig.\ref{ConC} è stratificato verticalmente e orizzontalmente.\\ 

Orizzontalmente, da sinistra a destra, il punteggio $u$ è stato aumentato. Se uno studente ha una risposta
tale che $u<0,4$  è casuale;  se $0,4 \leq u<0,7$ la risposta è media e  per un $u \geq 0,7$ la risposta  
è  capita come alta.  Verticalmente la concentrazione aumenta dal basso verso l'alto, affinché se $C<0,2$ le 
risposte degli studenti sono distribuiti in parti molto simili formano un istogramma quasi piatto tra tutte 
le opzioni, se  $0,2 \leq C<0,5$ la distribuzione delle risposte  è effettuata prevalentemente su due 
opzioni e se  $C \geq 0,5$ la distribuzione è predominante su  un'opzione.\\

Quindi, potrebbero esserci diverse combinazioni. Per esempio potrebbe presentarsi alta concentrazione 
con punteggio casuale, che significa molti studenti con un'idea sbagliata molto radicata; anche potrebbe 
presentarsi alta concentrazione  con punteggio elevato,  che significa molti studenti con un'idea corretta 
e stabile, ecc.

\section{Metodologia}
In questa sezione si presentano gli aspetti metodologici dello studio: caratteristiche della popolazione, 
riferimento teorico della fisica, definizione del MCF, strumenti di indagine e progettazione dell'esperimento.\\

\subsection{Popolazione}
L'esperimento è stato sviluppato durante il secondo semestre dell'anno  $2015$ in due università di 
Bogotà (Colombia). Da qui in avanti queste università   si chiameranno $X$ e $Y$, rispettivamente.  Gli 
studenti di $X$ studiavano ingegneria e quelli di $Y$, erano formati come insegnanti di chimica.\\
 
Nell'università $X$ l'esperimento è stato effettuato con due gruppi di un primo corso di fisica 
notturno, ogni gruppo aveva $13$ uomini e $3$ donne di età compresa tra $18$ e $40$ anni, la maggioranza 
degli studenti avevano un lavoro al giorno. Nell'università $Y$ l'esperimento è stato effettuato con solo 
gruppo di un primo corso di fisica nel giorno con $8$ uomini e $8$ donne con un'età  tra $16$ e $18$ anni, 
nessuno studente lavorava. In tutti i corsi avevano più o meno $20$ studenti, ma sono stati scelti solo $16$ per
 un corso, in base ai criteri di frequenza e interesse per l'apprendimento.

\subsection{Contesto fisico: caduta gravi nell'aria}

Si descrive il contesto fisico sul quale si ha fatto la trasposizione didattica e si ha definito il MCF.\\

Quando un corpo cade in un ambiente calmo, come l'aria in un ambiente controllato, tre forze  
agiscono su questo: il peso, la spinta e l'attrito con l'aria. La forza di resistenza dell'aria è meno 
complicata del peso e della spinta, perché la prima dipende dall'area trasversale del corpo,  del quadrato 
della  sua velocità,  della densità del mezzo e del coefficiente di trascinamento $c$, che a sua volta dipende 
dalla viscosità del mezzo $b$  e dalla  forma del corpo $R_e$  \cite{rise,EXV}; mentre le altre due forze sono 
costanti.\\

Come problema di riferimento si è studiato il problema di un corpo sferico che sta cadendo 
attraverso l'aria. La caduta degli altri corpi è stata paragonata a questo particolare problema. Per risolvere
questo problema il numero di Reynolds $R_e$ è molto importante, ed è stato definito come

%---
\begin{equation}
    R_e =\frac{\rho_m L v}{b},
\end{equation}
%--- 
qui $\rho_m$  è la densità del mezzo, $L$ la lunghezza della sezione trasversale del corpo,  $v$  la sua 
velocità e $b$  il coefficiente di viscosità del  mezzo. Il coefficiente di trascinamento $c$ si ottiene da 
esperimenti, in particolare per una sfera nell'intervallo $0<R_e<2 \times 10$ è collegato al numero di Reynolds
così 

%--------------
\begin{equation}\label{CRe}
    c=\frac{24}{R_e}+\frac{6}{1+\sqrt{R_e}}+0.4.
\end{equation}
%--------------
In queste condizioni un corpo sferico di raggio $r$ avrà un'accelerazione
%----
\begin{equation}\label{ace1}
    a=\left (1-\frac{\rho_m}{\rho_c}\right )g+\frac{3\rho_m c v^2}{8\rho_c r}.  
\end{equation}
%----

%---------
A causa della natura di $c$ nella eqn.(\ref{CRe}) non è possibile trovare una soluzione analitica generale 
della eqn.(\ref{ace1}), nel  caso particolare di corpi sferici e anche se è possibile trovare alcuna 
soluzione analitica si risolve numericamente  per comodità \footnote{È stato utilizzato  il metodo di 
Runge-Kutta quarto ordine \cite{EXV}} . Tuttavia, non si deve perdere di vista che si tratta di un caso particolare e in 
nessun modo si risolvono  tutti i casi in questo modo.\\

\begin{figure*}[ht] % puede tener problemas con [htbl!]
			\begin{minipage}[b]{0.5\linewidth} %Una minipágina que cubre la mitad de la página
				%\centering
					\includegraphics[width=9cm]{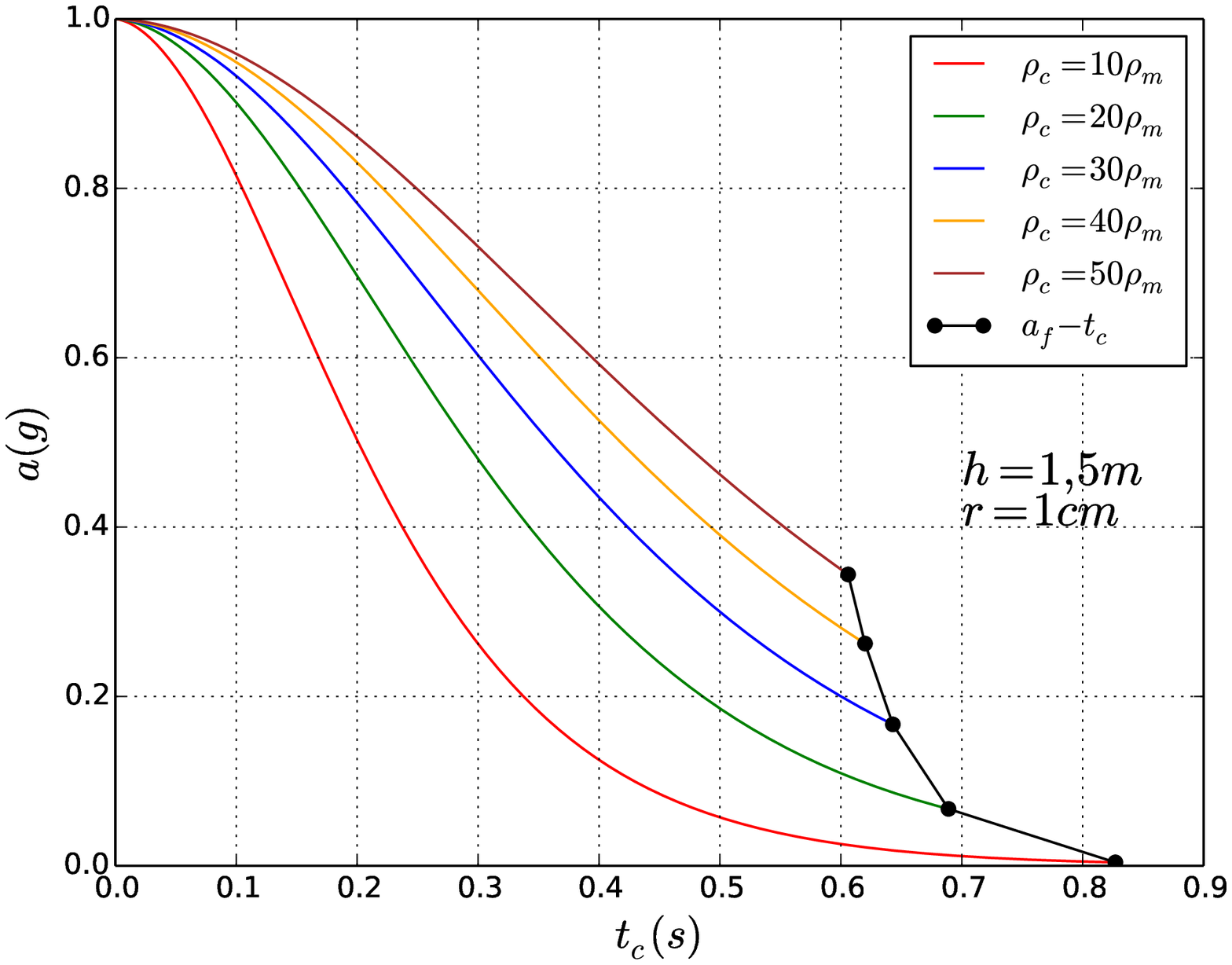}
					\caption{$a$ vs $t_c$ con $r$ costante.} \label{at1}
			\end{minipage}
					\hspace{0.0cm} % Si queremos tener un poco de espacio entre las dos figuras
			\begin{minipage}[b]{0.5\linewidth}
				%\centering
				\includegraphics[width=9cm]{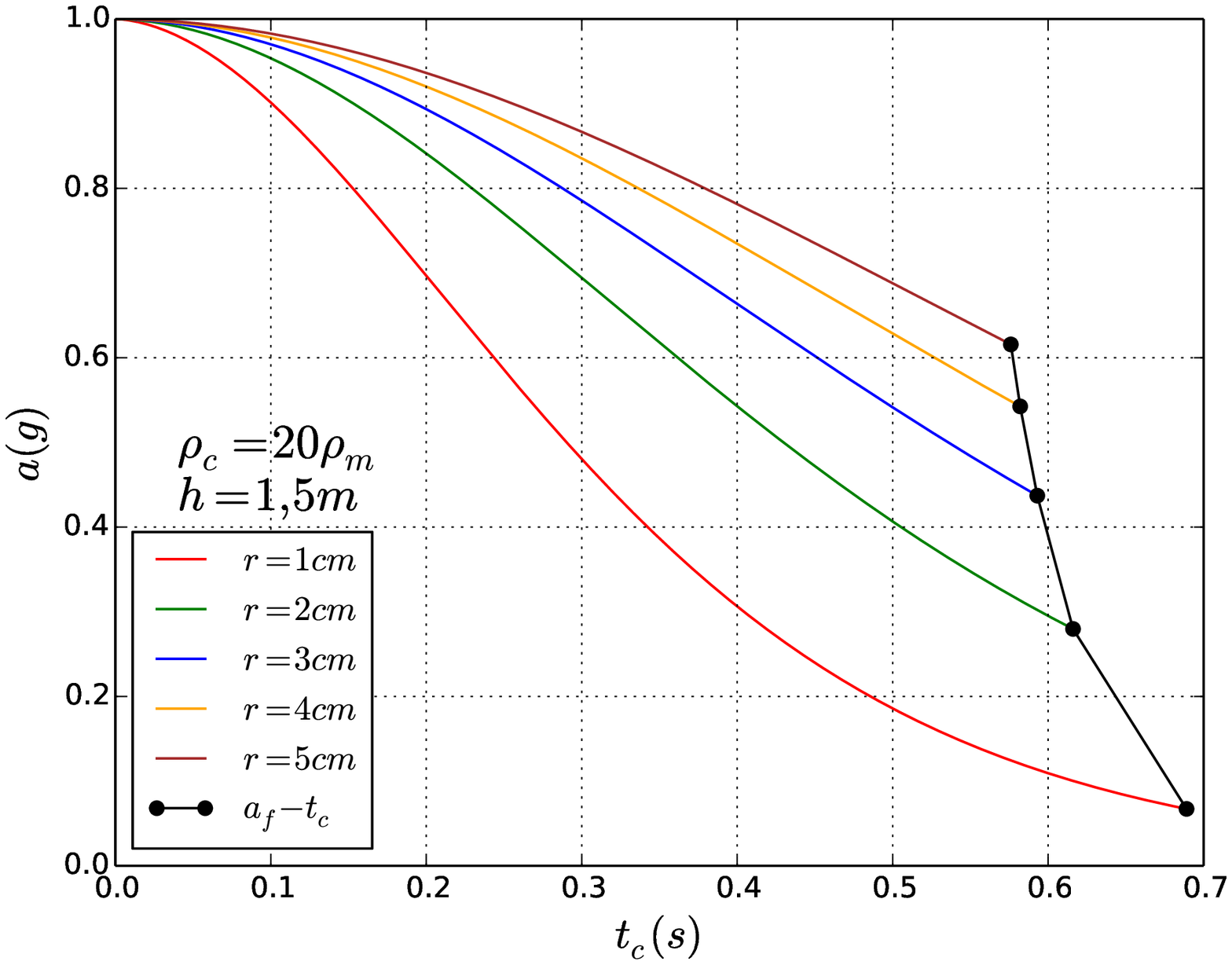}
				\caption{$a$ vs $t_c$ con $\rho_c$ costante.} \label{at2}
			\end{minipage}
		\end{figure*}

Quando due corpi sferici si fanno cadere simultaneamente dalla stessa altezza il valore dell'accelerazione del 
corpo ha un rapporto inverso (anche se non è proporzionale) con il tempo di caduta, cioè a maggiore intensità 
dell'accelerazione minore tempo di caduta. Questo si può vedere nei grafici delle Fig.\ref{at1} e Fig.\ref{at2},
 che sono state generate dal metodo numerico. 

%-----Doble gráfico--------EL * HACE QUE LA GRÁFICA O LA TABLA OCUPEN LAS DOS COLUMNAS
	
\subsection{Definizione del MCF}
Ragazzi, attenzione! 
Dice il professore nella sua classe. Dopo,  continua: 
ascoltatemi, se libero dalla stessa altezza e simultaneamente questi due corpi, quale arriva 
prima a terra?   Si è fatta ricerca sulla risposta a questa domanda in  diverse popolazioni di 
studenti e si sono trovate due idee sbagliate: gli studenti pensano che cadrà a terra prima il 
corpo più pesante oppure che i corpi cadano al momento \cite{LieB_3,H1,FCI}.\\

La prima risposta è identificata in questo lavoro come pensiero Aristotelico e la seconda come 
pensiero Galileiano. Sia il pensiero Aristotelico che il Galileiano,  non sono, in generale, d'accordo 
con i risultati sperimentali \footnote{Naturalmente i pensieri Aristotelici e Galileiani hanno  
molteplici sfumature, ma in questo lavoro si farà riferimento solo ai due aspetti menzionati prima: 
un corpo pesante impiega meno tempo per arrivare a terra che uno leggero oppure arrivano a terra in tempi 
uguali}.  In contrapposizione c'è la meccanica Newtoniana che spiega in modo coerente i fatti sulla 
caduta dei corpi,  che  sarà chiamata d'ora in poi,  pensiero Newtoniano.\\ 

Per definire il MCF si sono definiti tre attributi concettuali, ai pensieri Aristotelici 
$F_A$, Galileani $F_G$  e Newtoniani $F_N$. A ogni attributo 
concettuale sono associati $4$, dal più piccolo al più grande, livelli di discernimento 
denotati come $q_0$, $q_1$, $q_2$ e $q_3$; vedi tabella (\ref{tS}).
%---hasta aqui segunda corrección del P vito
%------------------
\begin{table}[htp]
\centering
\begin{tabular}{|c||c|c|c|}
 \hline  
 \textsf{} & $F_A$  & $F_G$  & $F_N$  \\ \hline  \hline
 $q_0$      &  &  &  \\ \hline  
 $q_1$      &  &  &  \\ \hline  
 $q_2$      &  &  &   \\ \hline 
 $q_3$      &  &  &  \\ \hline  
\end{tabular}
\caption{Definizione del MCF}\label{tS} 
\end{table}
%----------
\subsection{Strumento e procedura di indagine}

Per realizzare l'esperimento si è scritto  un piccolo testo sopra la teoria dei corpi in caduta, una 
guida didattica con tre esperimenti e un elenco di nove domande con risposta vera o falsa,  vedi 
l'appendice \textbf{A}.\\

L'esperimento di apprendimento  è stato strutturato nel seguente ciclo:
\begin{enumerate}
\item Il professore parla con i suoi studenti dell'esperimento ma non lo fa.  Dopo che  il professore avrà finito la sua spiegazione, farà una domanda e gli studenti
 avranno $5$ minuti per risponderla individualmente.    
\item Il professore farà gruppi di $2$, $3$ o $4$ studenti. Dopo che gli studenti faranno l'esperimento, risponderanno  la stessa domanda, ma in accordo con il
 gruppo. Il tempo per questa attività sarà vicino ai $10$ minuti.
\item Il professore farà una breve spiegazione basata su errori comuni e inviterà  alla discussione e riflessione attraverso le formulazioni e soluzioni  di domande.
\item Poi si torna al passo $1$, ma i gruppi di studenti saranno diversi dagli anteriori.
\end{enumerate}

%--------------
\section{Risultati}
\begin{figure}[ht] %[!htp]
	\begin{center}
	\includegraphics[scale=0.42]{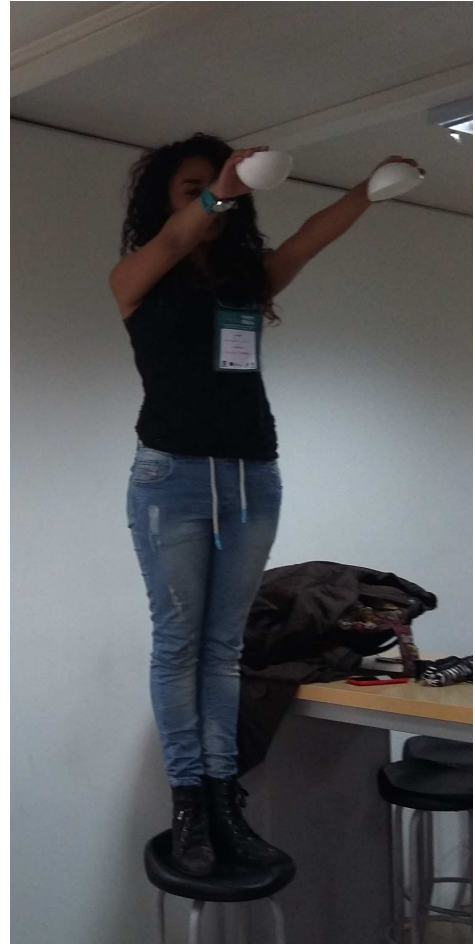}
	\caption{Una studentessa fa un esperimento sulla caduta di due corpi identici in massa, volume, forma, 
	         densità e area trasversale, ma con una faccia   diversa  al movimento. } 
	\label{kst}
	\end{center}
\end{figure}

I risultati delle risposte degli studenti sono stati organizzate attraverso tre digiti tra $0$ a $3$; il primo 
si riferisce  al livello del  pensiero Aristotelico; il secondo al livello del pensiero Galileiano e il terzo 
al livello del pensiero Newtoniano, vedi Fig.\ref{kst}. Così ogni pensiero corrisponde ai codici $030$, $300$, $333$, 
rispettivamente. Se lo studente non è identificato con questo codice significa che ha un tipo de 
pensiero ambiguo,  questa è la quarta opzione in una domanda con $4$ opzioni, vedi
tabella \ref{pregunta}.\\ 

 %----------grafica en otro lado
%----  

Nelle tabelle \ref{tab19}, \ref{tab13} e \ref{tabUD}  dell'appendice \textbf{B} si mostrano le risposte degli studenti. I 
numeri $1,2,3, \dots$ rappresentano ogni studente; $t_1,t_2,\dots$ rappresenta ciascuna delle 
lezioni; i numeri in ogni quadrato rappresentano le risposte degli studenti. 
%------Tabla pregunta equivalente
\begin{table}[htp]
\centering
\begin{tabular}{|c||c|c|c|}
 \hline  
 Opzione & Codice         & Pensiero         &   Modo           \\ \hline  \hline
 $A$   &  $333$   	      & Newtoniano       &   Corretto       \\ \hline  
 $B$   &  $300$           & Galileiano       &   Sbagliato   \\ \hline 
 $C$   &  $030$           & Aristotelico     &   Sbagliato   \\ \hline  
 $D$   &  Ogni altro      & Miscela          &   Sbagliato 	\\ \hline    
\end{tabular}
\caption{Domanda equivalente.}
\label{pregunta} 
\end{table}

%----------------
\subsection{Analisi del profitto $G_M$}

Per rendere conto del profitto si prende come punto di riferimento la prima volta che lo strumento è 
stato applicato, vale a dire $t_1$, questo è stato fatto con ogni gruppo  $gA$, $gB$ e $gC$, i cui 
risultati sono  mostrati nella  tabella \ref{tabGMEx}.

\begin{figure}[!htp] %[!htp]
	\begin{center}
	\includegraphics[scale=0.47]{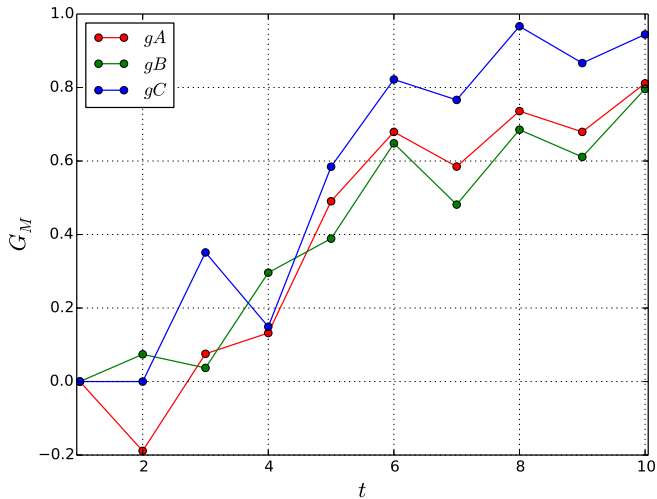}
	\caption{Curve di apprendimento. Il guadagno è stato valutato prendendo come punteggio
	   iniziale $k=1$ quello mostrato in to e ogni $t_k$ permette di valutare momento per momento il punteggio. }
	\label{excur}
	\end{center}
\end{figure}

%------Tabla de comparacion de ganancia de grupos experimentales
\begin{table}[htp]
\centering
\begin{tabular}{|c||c|c|c|c|}
 \hline  
 Grupo & $\epsilon_o$ & $\epsilon_f$ &$G_M$ & Livello di $G$ \\ \hline  \hline
 $gA$   &  $0,5$   	  & $0,9$        & 	$0,8$   & Elevato		\\ \hline  
 $gB$   &  $0,46$      & $0,89$       &  $0,8$ 	& Elevato	    \\ \hline  
 $gC$   &  $0,31$      & $0,97$       &  $0,96$ & Elevato		\\ \hline    
\end{tabular}
\caption{Profitto $G_M$.}\label{tabGMEx} 
\end{table}
%-----
L'evoluzione del  profitto attraverso il tempo si mostra nella  Fig.\ref{excur}. Qui si osserva come 
i gruppi $gA$ e $gB$ dell'università $X$ lo hanno evoluto in forme  simili, ma alla fine il gruppo $gA$ ha 
avuto più apprendimento che il gruppo $gB$. D'altra parte, il gruppo $gC$ dell'università $Y$ ha avuto 
più apprendimento che  gli altri due  gruppi, soprattutto verso la fine del processo. 
 %-------------------
%----  
\subsection{Analisi di concentrazione}

\begin{figure}[!htp]
\begin{center}
	\begin{tikzpicture}[domain=0:1,scale=6.2]
	\draw [-latex,black,line width=1.5pt] (-0.1cm,0cm) -- (1.1cm,0cm);%--Eje x----
	\draw [-latex,black,line width=1.5pt] (0cm,-0.1cm) -- (0cm,1.1cm);%--Eje y---
	\coordinate [label=below:\textcolor{black} {$u$}] (x) at  (1.12cm,0.05cm);%-x
	\coordinate [label=below:\textcolor{black} {$C$}] (x) at  (0.06cm,1.12cm);%-y
	\coordinate [label=below:\textcolor{black} {$1$}] (x) at  (1.0cm,0.0cm);%-x
	\coordinate [label=below:\textcolor{black} {$1$}] (x) at  (-0.05cm,1.05cm);%-y
	\coordinate [label=below:\textcolor{black} {$0,4$}] (x) at  (0.4cm,0.0cm);%-0,4
	\coordinate [label=below:\textcolor{black} {$0,7$}] (x) at  (0.7cm,0.0cm);%-0,4
	\coordinate [label=below:\textcolor{black} {$0,2$}] (x) at  (-0.1cm,0.25cm);%-0,2
	\coordinate [label=below:\textcolor{black} {$0,5$}] (x) at  (-0.1cm,0.55cm);%-0,2
	
     \draw [black,line width=0.5pt] (0cm,1.0cm) -- (1cm,1cm);%----
     \draw [black,line width=0.5pt] (1cm,1.0cm) -- (1cm,0cm);%----
     \draw [black,dashed,line width=0.5pt] (0cm,0.2cm) -- (0.7cm,0.2cm);%----
     \draw [black,dashed,line width=0.5pt] (0cm,0.5cm) -- (1cm,0.5cm);%----
     \draw [black,dashed,line width=0.5pt] (0.4cm,0.0cm) -- (0.4cm,1.0cm);%----
     \draw [black,dashed,line width=0.5pt] (0.7cm,0.0cm) -- (0.7cm,1.0cm);%----   
     %-------------Funciones m=4--------------------
     \draw[color=black,line width=0.5pt]   plot ({\x},{2.0*sqrt(1-2.0*\x+2.0*\x*\x)-1.0}); %\x-(\x*\x)/2
	\draw[color=black,line width=0.5pt]   plot ({\x},{2.0*sqrt(1.0/3.0-2.0*\x/3.0+4.0*\x*\x/3.0)-1.0});
	%------------Evolución
	\draw [-latex,red,line width=1.2pt] (0.0cm,0.768cm) -- (0.375cm,0.458cm);%---Grupo 19uc
	\draw [-latex,green,line width=1.2pt] (0.0cm,0.768cm) -- (0.25cm,0.58cm);%---Grupo 13uc
	\draw [-latex,blue,line width=1.2pt] (0.0cm,0.768cm) -- (0.69cm,0.51cm);%---Grupo C ud
	%------------Marcas
	\draw [red,line width=1.2pt]   (0.8cm,0.4cm) -- (0.9cm,0.4cm);%---Grupo 19uc
	      \coordinate [label=below:\textcolor{black} {$gA$}] (x) at  (0.95cm,0.46cm);
	\draw [green,line width=1.2pt] (0.8cm,0.3cm) -- (0.9cm,0.3cm);%---Grupo 13uc
	    \coordinate [label=below:\textcolor{black} {$gB$}] (x) at  (0.95cm,0.36cm);
	\draw [blue,line width=1.2pt]  (0.8cm,0.2cm) -- (0.9cm,0.2cm);%---Grupo C ud
		\coordinate [label=below:\textcolor{black} {$gC$}] (x) at  (0.95cm,0.25cm);
	\end{tikzpicture}
\caption{Diagramma di concentrazione contro prestazioni.}
\label{Condis}
\end{center}
\end{figure}
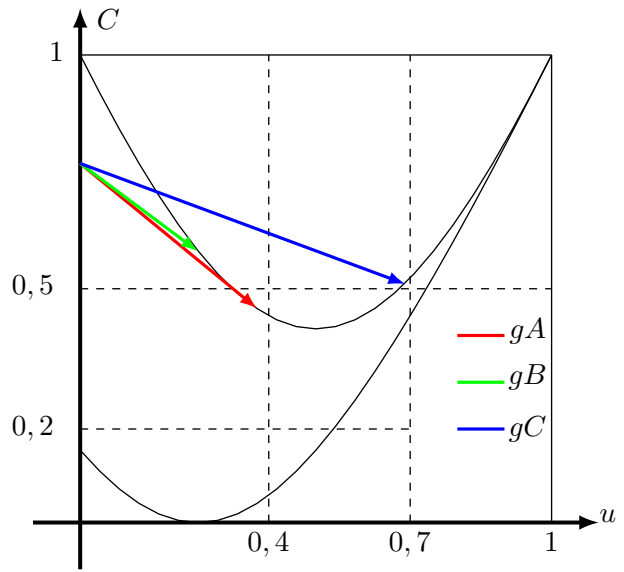

La Fig.\ref{Condis} mostra i risultati della concentrazione contro il punteggio.  Si osserva che tutti i 
gruppi hanno lo stesso punteggio e concentrazione iniziale, che corrisponde a un'alta concentrazione in 
un'idea errata,  mentre lo stato finale è diverso per ogni gruppo.  Il gruppo $gA$ nel suo punto finale
mostra basso punteggio con la presenza di due opzioni, dove la maggioranza degli studenti hanno pensiero 
Newtoniano e la minoranza hanno una miscela degli altri tre pensieri.  Il gruppo $gB$ ha più punteggio che 
il gruppo $gA$, ma ha meno concentrazione che il gruppo $gA$, qui ci sono solo pochi  studenti con pensiero 
Newtoniano. Il gruppo $gC$ ha ottenuto un punteggio superiore ed un'alta concentrazione, qui la maggior 
parte degli studenti hanno un pensiero Newtoniano.\\
%--------------------

%-------\twocolumn[\section{This is a very wide section heading spanning over two columns}]

\section{Discussione}

In questa investigazione si è potuto risolvere, in parte, il problema dell'influenza che gli studenti 
avanzati fanno sopra gli studenti medi nella metodologia PI. Questo si è fatto attraverso una rotazione 
in ogni interazione tra studenti in modo che non sempre si formino gli stessi gruppi di studenti. Tuttavia 
non è stato sempre possibile mantenere gruppi veramente diversi,  si è anche trovato un apprendimento più
alto e piú lineare che in altre ricerche \cite{Quepasa,Quepasa2}.\\  

Le domande dallo strumento di inchiesta non sono tutte teoriche, hanno anche domande sperimentali, 
sebbene non tutte le domande hanno una spiegazione a partire dal referente teorico, si è indotta un'ipotesi 
tacita che permette differenziare i vari tipi di pensiero.\\

La linearità trovata, senza picchi e interferenze, nelle curve di profitto è dovuta alla frequenza con 
cui le attività  sono state applicate;  non c'era tempo per gli studenti di dimenticare ciò che avevano 
imparato. In altri studi come quelli realizzati da Eleanor C. Sayre e Andrew F. Heckler, gli studenti 
non solo si sono dimenticati, ma si sono confusi e per questo motivo le loro curve di profitto mostrano 
picchi e cali e interferenze \cite{Quepasa2}.

\section{Conclusioni}
A partire da principi cognitivi ben consolidati nella letteratura è stato formulato un modello di 
apprendimento sperimentale di un MCF basata su PI. Tale modello è stato confrontato con l'esperienza 
di apprendimento in due università e tre gruppi di studenti, attraverso le curve di apprendimento si 
osserva un'ipotesi positiva sopra un migliore apprendimento. Il lavoro ha dimostrato che se si fa un 
controllo dell'influenza degli studenti  di livello alto e si fanno esperimenti i risultati  di 
apprendimento sono elevati, con $G_M \approx 0,96$. 

\section*{Ringraziamenti}
L'autore ringrazia  alla facoltà di ingegneria e scienze di base dell'università 
Central per il supporto nello svolgimento di questo lavoro nell'ambito del corso di
laurea magistrale in modellistica e simulazione. Ringrazia anche: ai suoi colleghi Fernanda 
Santana, Orlando Organista e Luis Barbosa per aver servito come comitato di esperti nella 
valutazione degli strumenti di indagine e per tutti i suggerimenti dati nelle discussioni che 
hanno arricchito tale argomento; ai professori Vito Ottati e Guillermo Avendaño per fare la revisione 
del manoscritto in italiano ed in inglese, rispettivamente;  all'insegnante Karen Gonzalez 
(che era studente quando è stata condotta questa ricerca) per autorizzare la pubblicazione 
della sua fotografia sca\-tta\-ta durante la ricerca.

\renewcommand{\refname}{Bibliografia}

\include{AQ}

\include{tableR}

\end{document}

%% file: AQ.tex
\onecolumn
\section*{Appendice $A$: Strumenti di indagine } \label{app:guias}

Gli strumenti di indagine sono stati taggati come esperimenti  $\alpha, \beta, \gamma, \delta$, e $\epsilon$, i quali si sono presentati come segue: 

\emph{Nel seguente testo si descrivono tre esperimenti con i suoi risultati, d'accordo con loro segnare vero \textbf{V} oppure falso \textbf{F} per 
ciascuna delle affermazioni scritte sotto.}  

\section*{\textbf{A-$1.$} Esperimenti $\alpha$}
\subsection*{Esperimento $\alpha_1$}

In classe il professore prende un foglio di carta pulito lo taglia in due parti leggermente disuguali, prende il più grande e fa una palla 
accartocciando la carta. Dopo prende in una mano la palla e nell'altra il pezzo di carta senza accartocciare. Dalla stessa altezza 
li lascia cadere tutte e due allo stesso tempo. Si osserva che il primo che arriva a terra e la palla.
\begin{enumerate}
\item Il risultato avviene perché i corpi più pesanti cadono sempre prima di quelli leggeri.  \textbf{V}  \textbf{F}

\item Il risultato è d'accordo con la legge di Galileo sulla caduta dei gravi poiché i pezzi di carta cadono con la stessa accelerazione 
      indipendentemente dal loro peso eppure come hanno differenti dimensioni non cadono allo stesso tempo.   \textbf{V}  \textbf{F}

\item Tutti i corpi con la stessa forma cadono sempre allo stesso tempo ma come in questo esperimento i pezzi di carta non hanno 
      la stessa forma quindi non cadono allo stesso tempo. \textbf{V}  \textbf{F}
\end{enumerate}

\subsection*{Esperimento $\alpha_2$}

Adesso il professore raccoglie i pezzi di carta dal suolo e scartoccia il pezzo accartocciato. Dopo accartoccia il pezzo liscio formando una
palla, successivamente ne prende ognuno in ogni mano e dalla stessa altezza li lascia cadere. Ancora una volta, il primo che tocca terra è 
la palla. 

\begin{enumerate}
\item I risultati dell'esperimento sono d'accordo con la legge di Galileo sulla caduta dei gravi poiché i pezzi di carta cadono con la stessa 
      accelerazione indipendente dal loro peso eppure come hanno differenti dimensioni non cadono simultaneamente. \textbf{V}  \textbf{F}
\item L'esperimento dimostra che un corpo leggero potrebbe arrivare a terra, prima di uno più pesante. \textbf{V}  \textbf{F}
\item Siccome corpi della stessa forma cadono sempre allo stesso tempo ma in questo esperimento i pezzi di carta non hanno la stessa forma quindi
    non cadono allo stesso tempo. \textbf{V}  \textbf{F}
\end{enumerate}

\subsection*{Esperimento $\alpha_3$}

\begin{enumerate}
\item Alla fine il professore raccoglie i pezzi di carta, li accartoccia formando due palle, ne prende una in ogni mano e le lascia cadere 
dalla stessa altezza. Si osserva che arrivano a terra allo stesso tempo. 
\item Quello che si osserva è d'accordo con la legge di Galileo sulla caduta dei gravi. \textbf{V}  \textbf{F}
\item D'accordo con l'esperimento è possibile che due corpi con peso diverso cadano allo stesso tempo.\textbf{V}  \textbf{F}
\item Siccome corpi con la stessa forma cadono sempre allo stesso tempo e in questo esperimento i pezzi di carta  hanno la stessa forma, cadono 
allo stesso tempo. \textbf{V}  \textbf{F}
\end{enumerate}

%--------beta--
\section*{\textbf{A-$2.$} Esperimenti $\beta$}

\subsection*{Esperimento $\beta_1$}
Il professore prende con una mano una moneta e con l'altra un pezzettino di carta più piccolo di essa. Li lascia cadere 
da un'altezza di $50cm$. Si osserva che cade prima la moneta. 
\begin{enumerate}
\item Il risultato avviene perché i corpi più pesanti cadono sempre prima di quelli leggeri. \textbf{V}  \textbf{F}
\item Si trovano questi risultati dato che corpi di diversa forma non cadono mai allo stesso tempo. Questo soltanto 
     succederebbe nel caso in cui i corpi cadessero nel vuoto così come afferma la legge di Galileo sulla caduta dei gravi. \textbf{V}  \textbf{F}
\item Il risultato si conferma giacché corpi di forma diversa non cadono mai simultaneamente. \textbf{V}  \textbf{F}
\end{enumerate}

%------
\subsection*{Esperimento $\beta_2$}

Adesso il professore prende il pezzettino di carta e lo mette sopra la moneta, li lascia cadere da un'altezza di $50cm$. Si osserva 
che non si staccano prima di arrivare a terra. 
\begin{enumerate}
\item Il risultato è d'accordo con la legge di Galileo poiché tutti i corpi cadono allo stesso tempo indipendentemente dal loro peso. \textbf{V}  \textbf{F}
\item Il risultato era da aspettarselo dato che i corpi con forma diversa non cadono mai simultaneamente ma come sono attaccati si
 può capire che sono un solo corpo e per questo cadono allo stesso tempo. \textbf{V}  \textbf{F}
\item Il risultato dimostra che due corpi con peso molto diverso potrebbero cadere allo stesso tempo. \textbf{V}  \textbf{F}
\end{enumerate}

%------
\subsection*{Esperimento $\beta_3$}

Alla fine il professore prende la moneta e la mette sotto il pezzettino di carta, li lascia cadere da 1m di altezza. Si osserva che subito 
si staccano e la moneta arriva prima a terra. 
\begin{enumerate}
\item Questo succede perché corpi di diversa forma non cadono  mai allo stesso tempo.\textbf{V}  \textbf{F}
\item Questo dimostra che in alcuni casi un corpo pesante può cadere prima di uno più leggero. \textbf{V}  \textbf{F}
\item L'esperimento non si può spiegare con la legge di Galileo sulla caduta dei gravi poiché solo si usa quando l'aria non influisce 
notevolmente nella loro accelerazione. \textbf{V}  \textbf{F}
\end{enumerate}

%----------Experimento gamma
\section*{\textbf{A-$4.$} Esperimenti $\gamma$}

\subsection*{Esperimento $\gamma_1$}

Il professore usa due palloncini identici, un filo e una biglia. Prima di tutto gonfia i palloncini uno più grande dell'altro, 
ne prende uno in ogni mano e li lascia cadere dalla stessa altezza. Arriva a terra prima il più piccolo. 
\begin{enumerate}
\item Siccome i palloncini non hanno le stesse dimensioni non cadono allo stesso tempo, altrimenti cadrebbero simultaneamente. \textbf{V}  \textbf{F}
\item Questo dimostra che in alcuni casi un corpo leggero può cadere prima di uno più pesante. \textbf{V}  \textbf{F}
\item Nella legge di Galileo sulla caduta dei gravi tutti i corpi indipendentemente dal loro peso cadono con la stessa 
  velocità, dato che i palloncini non cadono al contempo significa che hanno diverso peso. \textbf{V}  \textbf{F}
\end{enumerate}

%------
\subsection*{Esperimento $\gamma_2$}

Ora il professore mette la biglia di vetro in uno dei palloncini, li lascia cadere dalla stessa altezza e osserva che cade prima 
il palloncino senza biglia.
\begin{enumerate}

\item Per ottenere questo risultato è necessario che i palloncini abbiano la stessa forma e le stesse dimensioni poiché tutti i corpi 
      che hanno queste caratteristiche cadono allo stesso tempo. \textbf{V}  \textbf{F}
\item Se il palloncino con la biglia si gonfia di più è possibile che succeda questo giacché il corpo più leggero può cadere prima. \textbf{V}  \textbf{F}
\item D'accordo con Galileo tutti i corpi cadono allo stesso tempo indipendentemente dal loro peso quando non c'è l'aria ma come questo 
non è il caso, cade prima il più grande.\textbf{V}  \textbf{F}
\end{enumerate}

%------
\subsection*{Esperimento $\gamma_3$}

Adesso il professore prende i palloncini, ne gonfia uno più dell'altro e li lega con un filo molto leggero, li lascia cadere simultaneamente 
dalla stessa altezza. Dopo domanda cosa succederà. 
\begin{enumerate}
\item Dovuto alla legge di Galileo sulla caduta  dei gravi i palloncini cadono lentamente e il filo tra loro non si allunga. \textbf{V}  \textbf{F}
\item Siccome i palloncini hanno la stessa forma ma dimensioni diverse,  il palloncino più grande trascina a quello più piccolo e si estende il
 filo completamente. \textbf{V}  \textbf{F}
\item Dato che l'aria influisce di più sul palloncino più pesante, il palloncino leggero lo trascina ed estende il filo completamente. \textbf{V}  \textbf{F}
\end{enumerate}

%-----------Experimento delt
\section*{\textbf{A-$3.$} Esperimenti $\delta$}
     
\subsection*{Esperimento $\delta_1$}

In questa occasione il professore prende una sfera di polistirolo e la taglia a metà, ne prende una in ogni mano e le lascia cadere dalla stessa 
altezza con la parte curva di fronte al suolo. Si osserva che le sfere toccano terra allo stesso tempo.    
\begin{enumerate}
\item Il risultato è d'accordo con la legge di Galileo dato che i corpi cadono simultaneamente indipendentemente dal loro peso e il mezzo in cui 
      cadono. \textbf{V}  \textbf{F}
\item Il risultato dell'esperimento è d'accordo con la legge di Galileo sulla caduta dei gravi la quale dice che corpi della stessa forma, peso
       e dimensioni cadono sempre allo stesso tempo. \textbf{V}  \textbf{F}
\item Il risultato era di aspettarselo giacché si sà che l'unica maniera nella quale due corpi  cadono al contempo in presenza dell'aria è che 
      abbiano lo stesso peso. \textbf{V}  \textbf{F}

\end{enumerate}

%----
\subsection*{Esperimento $\delta_2$}

Adesso il professore lascia cadere della stessa altezza e simultaneamente le sfere, ma questa volta una con la parte curva di fronte al suolo e l'altra 
con la parte piana nella parte opposta. Arriva prima la sfera che si lascia cadere con la parte curva. 
\begin{enumerate}
\item La legge di Galileo sulla caduta dei gravi afferma che senza l'aria tutti i corpi cadono con la stessa accelerazione indipendente dal loro peso, 
     visto che c'è l'aria questa legge non si applica e per tanto le sfere non cadono allo stesso tempo. \textbf{V}  \textbf{F}
\item L'esperimento dimostra che due corpi con la stessa forma, le stesse dimensioni e lo stesso peso possono non cadere allo stesso tempo. \textbf{V}  \textbf{F}
\item L'esperimento dimostra che due corpi dello stesso peso non necessariamente arrivino a terra allo stesso tempo. \textbf{V}  \textbf{F}
\end{enumerate}

%------
\subsection*{Esperimento $\delta_3$}

Alla fine il professore prende una sfera e mette alcune monete nella parte piana. Lascia cadere la sfera senza monete con la parte curva 
di fronte al suolo mentre la sfera con le monete la lascia cadere con la parte piana di fronte al suolo dalla stessa altezza. Cade prima la 
sfera con le monete. 
\begin{enumerate}
\item  L'esperimento dimostra che, sempre che due corpi si lasciano cadere allo stesso tempo, arriva a terra prima il più pesante. \textbf{V}  \textbf{F}
\item  Il risultato è d'accordo con la legge di Galileo sulla caduta dei gravi giacché i corpi più pesanti arrivano a terra prima per natura, 
      e siccome la sfera con le monete pesa di più, questo soddisfa la legge. \textbf{V}  \textbf{F}
\item  L'esperimento dimostra che, due corpi con la stessa forma e le stesse dimensioni ma con diverso peso non cadono mai 
     contemporaneamente. \textbf{V}  \textbf{F}
\end{enumerate}

%-----------Experimento delepsilon
\section*{\textbf{A-$5.$} Esperimenti $\epsilon$}

\subsection*{Esperimento $\epsilon_1$}

Per questo esperimento il professore utilizza una sfera di polistirolo e una biglia, li lascia cadere da 15cm di altezza. Si osserva che cadono 
allo stesso istante.

\begin{enumerate}
 \item L'esperimento dimostra che due corpi di peso diverso possono arrivare a terra simultaneamente. \textbf{V}  \textbf{F}
 \item Il risultato  è d'accordo con la legge di Galileo sulla caduta dei gravi dato che quando i corpi cadono hanno la stessa accelerazione
       senza importare il loro peso, per questo arrivano a terra al contempo. \textbf{V}  \textbf{F}
 \item L'esperimento dimostra che due corpi con la stessa forma cadono sempre contemporaneamente. \textbf{V}  \textbf{F}
\end{enumerate}

%------
\subsection*{Esperimento $\epsilon_2$}

Adesso il professore lascia cadere allo stesso tempo la sfera e la biglia da una altezza di $2,5m$ approssimativamente dal tetto dell'aula. Si 
osserva che la biglia arriva a terra prima.
\begin{enumerate}
\item  Il risultato di questo esperimento non si può spiegare con la legge di Galileo poiché l'aria influisce in modo significativo sull'accelerazione 
       della sfera. \textbf{V}  \textbf{F}
\item L'esperimento dimostra che secondo sia l'altezza dalla quale si lasciano cadere i corpi possono arrivare a terra al contempo oppure no. Inoltre 
      si può inferire che questo fatto fisico è indipendente dal peso di ogni corpo. \textbf{V}  \textbf{F}
\item L'esperimento dimostra che è possibile che corpi della stessa forma non cadano simultaneamente. \textbf{V}  \textbf{F}
\end{enumerate}
%------
\subsection*{Esperimento $\epsilon_3$}
Infine, il professore lascia cadere simultaneamente la biglia e la sfera ma ognuna da una altezza diversa. Si osserva che arrivano a terra 
allo stesso tempo.

\begin{enumerate}
\item Questo risultato è possibile se si lascia cadere la biglia da una maggior altezza poiché tutti i corpi pesanti arrivano a terra prima di 
     quelli leggeri.\textbf{V}  \textbf{F}
\item Questo risultato è possibile se si lascia  cadere la biglia da una maggior altezza dato che secondo la legge di Galileo sulla caduta dei 
      gravi tutti i corpi indipendentemente dal mezzo cadono allo stesso tempo. \textbf{V}  \textbf{F}
\item Questo risultato è possibile se si lascia cadere la biglia da una maggior altezza perché cosí avrà una maggior accelerazione senza importare 
      che abbiano la stessa forma. \textbf{V}  \textbf{F}
\end{enumerate}

%% file: tableR.tex
\section*{Appendice $B$: Risultati sperimentali}
%---------tabla grupo--19---A
\begin{table*}[htp]
\centering
 \resizebox{.5\textwidth}{!}{% <------ Don't forget this %
\begin{tabular}{|c||c|c|c|c|c|c|c|c|c|c|}
 \hline  
       & $t_1$& $t_2$& $t_3$& $t_4$& $t_5$& $t_6$& $t_7$& $t_8$& $t_9$& $t_{10}$ \\ \hline  \hline
    $1$  & 111	& 300	& 213	& 212	& 322	& 332	& 231	& 333	& 233	& 333	\\ \hline 
	$2$  & 211	& 210	& 121	& 211	& 211	& 332	& 232	& 332	& 233	& 333	\\ \hline	
	$3$  & 122	& 222	& 222	& 332	& 222	& 332	& 332	& 332	& 331	& 332	\\ \hline
	$4$  & 303	& 200	& 212	& 211	& 332	& 332	& 332	& 332	& 332	& 333	\\ \hline
	$5$  & 200	& 300	& 211	& 212	& 332	& 332	& 232	& 332	& 321	& 332	\\ \hline
	$6$  & 131	& 210	& 201	& 211	& 323	& 332	& 233	& 332	& 332	& 333	\\ \hline
	$7$  & 203	& 200	& 111	& 211	& 222	& 321	& 332	& 123	& 321	& 331	\\ \hline
	$8$  & 211	& 300	& 212	& 221	& 221	& 321	& 321	& 223	& 333	& 333	\\ \hline
	$9$  & 121	& 100	& 121	& 212	& 332	& 332	& 311	& 333	& 331	& 331	\\ \hline
	$10$ & 300	& 300	& 311	& 211	& 332	& 332	& 332	& 223	& 333	& 333	\\ \hline
	$11$ & 020 	& 100	& 220	& 211	& 22	& 112	& 111	& 123	& 233	& 331	\\ \hline
	$12$ & 202	& 222	& 111	& 211	& 12	& 332	& 332	& 332	& 332	& 332	\\ \hline
	$13$ & 213	& 210	& 210	& 211	& 233	& 332	& 332	& 332	& 231	& 331	\\ \hline
	$14$ & 300	& 300	& 123	& 212	& 332	& 332	& 231	& 332	& 231	& 332	\\ \hline
	$15$ & 131	& 102	& 122	& 221	& 222	& 332	& 332	& 333	& 323	& 323	\\ \hline
	$16$ & 322	& 323	& 332	& 332	& 332	& 332	& 333	& 223	& 322	& 332	\\ \hline  
\end{tabular}
}
\caption{Risultati sperimentali dei gruppo $A$.}\label{tab19} 
\end{table*}
%---------tabla grupo--13---B
\begin{table}[htp]
\centering
\resizebox{.5\textwidth}{!}{% <------ Don't forget this %
\begin{tabular}{|c||c|c|c|c|c|c|c|c|c|c|}
\hline  
       & $t_1$& $t_2$& $t_3$& $t_4$& $t_5$& $t_6$& $t_7$& $t_8$& $t_9$& $t_{10}$ \\ \hline  \hline
    $1$& 0& 	0  & 	200& 	223& 	23& 	332& 	322& 	333& 	322& 	333 \\ \hline 
	$2$& 300& 	300& 	212& 	122& 	333& 	333& 	231& 	231& 	332& 	332 \\ \hline 
	$3$& 223& 	303& 	121& 	233& 	101& 	333& 	231& 	332& 	330& 	331 \\ \hline 
	$4$& 0& 	0  & 	121& 	233& 	212& 	332& 	231& 	231& 	333& 	331 \\ \hline 
	$5$& 300& 	300& 	222& 	233& 	233& 	323& 	323& 	232& 	321& 	332 \\ \hline 
	$6$& 313& 	312& 	320& 	212& 	302& 	321& 	121& 	232& 	333& 	332 \\ \hline 
	$7$& 313& 	323& 	223& 	223& 	332& 	332& 	331& 	331& 	332& 	333 \\ \hline 
	$8$& 323& 	323& 	213& 	223& 	211& 	321& 	322& 	333& 	333& 	333 \\ \hline 
	$9$& 120& 	303& 	221& 	122& 	332& 	332& 	332& 	232& 	232& 	332 \\ \hline 
	$10$& 330& 	300& 	211& 	111& 	323& 	222& 	331& 	232& 	323& 	332 \\ \hline 
	$11$& 10& 	323& 	211& 	212& 	322& 	332& 	232& 	331& 	233& 	333 \\ \hline 
	$12$& 200& 	300& 	121& 	111& 	121& 	222& 	221& 	332&    323& 	322 \\ \hline 
	$13$& 301& 	312& 	20 &    223& 	221& 	332& 	322& 	333& 	332& 	332 \\ \hline 
	$14$& 313& 	300& 	123 &   13& 	212& 	102& 	202& 	232& 	123& 	332 \\ \hline 
	$15$& 103& 	303& 	101	&   212& 	213& 	323& 	221& 	231& 	210& 	331 \\ \hline 
	$16$& 302& 	300& 	112 & 	212& 	233& 	332& 	133& 	332& 	231& 	332 \\ \hline 
\end{tabular}
}
\caption{Risultati sperimentali dei gruppo $B$.}\label{tab13} 
\end{table}
%---------tabla grupo--UD---C
\begin{table}[htp]
\centering
\resizebox{.5\textwidth}{!}{% <------ Don't forget this %
\begin{tabular}{|c||c|c|c|c|c|c|c|c|c|c|}
\hline  
       & $t_1$& $t_2$& $t_3$& $t_4$& $t_5$& $t_6$& $t_7$& $t_8$& $t_9$& $t_{10}$ \\ \hline  \hline
	$1$   &203	& 300	& 213	& 212	& 0		& 332	& 332	& 332	& 322	& 333	\\ \hline
	$2$   &332	& 130	& 0		& 0		& 221	& 332	& 322	& 333	& 333	& 333	\\ \hline
	$3$   &0	& 0		& 101	& 201	& 332	& 332	& 332	& 332	& 323	& 333	\\ \hline
	$4$   &202	& 203	& 222	& 223	& 323	& 332	& 333	& 333	& 333	& 333	\\ \hline
	$5$   &102	& 203	& 222	& 222	& 332	& 332	& 232	& 333	& 332	& 332	\\ \hline
	$6$   &0	& 0		& 0		& 212	& 221	& 332	& 331	& 333	& 313	& 333	\\ \hline
	$7$   &300	& 300	& 111	& 212	& 120	& 332	& 322	& 333	& 333	& 333	\\ \hline
	$8$   &110	& 200	& 211	& 212	& 321	& 332	& 131	& 333	& 331	& 332	\\ \hline
	$9$   &110	& 200	& 203	& 212	& 333	& 332	& 232	& 333	& 333	& 333	\\ \hline
	$10$   &300	& 100	& 213	& 212	& 322	& 332	& 333	& 333	& 333	& 333	\\ \hline
	$11$   &211	& 303	& 232	& 223	& 232	& 332	& 322	& 333	& 332	& 332	\\ \hline
	$12$   &1	& 303	& 222	& 212	& 332	& 332	& 332	& 333	& 331	& 332	\\ \hline
	$13$   &120	& 100	& 213	& 212	& 333	& 332	& 333	& 333	& 333	& 332	\\ \hline
	$14$   &31	& 130	& 112	& 223	& 122	& 332	& 332	& 333	& 333	& 333	\\ \hline
	$15$   &20	& 200	& 223	& 201	& 312	& 332	& 333	& 333	& 333	& 333	\\ \hline
	$16$   &0	& 0		& 331	& 201	& 213	& 332	& 332	& 332	& 332	& 333	\\ \hline	
\end{tabular}
}
\caption{Risultati sperimentali dei gruppo $C$.}\label{tabUD} 
\end{table}
%----------